\documentclass[floatfix]{revtex4}
\usepackage[dvips]{graphicx}
\usepackage{epsfig}

\usepackage[cp1251]{inputenc}
\usepackage[english]{babel}
\usepackage{color}
\usepackage{bm}

\usepackage[T1]{fontenc}
\usepackage{xcolor}
\usepackage{lmodern}
\usepackage{listings}

\begin{document}






\title{Non-Newtonian rheology in twist-bend nematic liquid crystals}

\author{E. I. Kats}  
\affiliation{Landau Institute for Theoretical Physics, RAS, \\
142432, Chernogolovka, Moscow region, Russia.}


\begin{abstract}
 In this work we present a simple qualitative model to describe
shear rheological behavior of the twist-bend nematic liquid crystals ($N_{TB}$).
We find that at relatively low shear rate (${\dot \gamma } \leq {\dot \gamma}_{c1}$) the stress
tensor $\sigma $ created by this shear strain, scales as $\sigma \propto {\dot \gamma }^{1/2}$. Thus the effective viscosity decreases with the shear rate ($\eta \propto {\dot \gamma }^{-1/2}$) manifesting so-called shear-thinning phenomenon. At intermediate shear rate ${\dot \gamma }_{c1} \leq {\dot \gamma} \leq {\dot \gamma }_{c2}$, $\sigma $ 
is almost independent of ${\dot \gamma }$ (a sort of plateau), and at large shear rate (${\dot \gamma } \geq 
{\dot \gamma}_{c2}$), $\sigma \propto {\dot \gamma }$, and it looks like as Newtonian rheology. Within our theory
the critical values
of the shear rate scales as ${\dot \gamma }_{c1}\propto ({\tilde {\eta }_2^0}/{\tilde {\eta }_3^0})^2$,
and ${\dot \gamma}_{c2} \propto ({\tilde {\eta }_2^0}/{\tilde {\eta }_3^0})^4$ respectively. Here
${\tilde \eta }_2^0$ and ${\tilde \eta }_3^0$ are bare coarse grained shear viscosity coefficients 
of the effective smectics equivalent to the $N_{TB}$ phase at large scales.
The results of our work are in the agreement with recent experimental studies.

\end{abstract}



\maketitle

{\bf{Background.}}
${\, }$

A number of exciting and relatively recent publications (about ten years ago, compared to more than 100 years of the discovery of the classical liquid crystals) report on a discovery of a new type of equilibrium liquid crystals, termed twist-bend nematics, $N_{TB}$ (see the papers \cite{HS09,PN10,CD11,BK13,MD14,SV20}). The discovery of $N_{TB}$ nematics 
opened ''Pandora box'' with new kinds of modulated liquid crystals (see very influential pioneering works and a few
review papers \cite{KA12} - \cite{SC20}).
Naturally (as it was the case in great geographical discoveries of 15-th - 17-th centuries)
after the first step devoted mainly to observations and structural identifications of new liquid crystals, the interest
moves to investigations and exploring of physical properties of these new phases. Since then the $N_{TB}$, and other modulated nematics
are becoming one of the hottest topics in physics of liquid crystals.

To start a few words about structural features of $N_{TB}$ nematics is in order here.
Especially surprising is the fact that the twist-bend nematics, $N_{TB}$) exhibit helical (chiral) orientational 
ordering despite being formed from achiral molecules. For comparison, there exist also chiral cholesteric phases locally equivalent to nematics but possessing simple (orthogonal) helical structures with pitches in a few $\mu m$ range. The cholesteric structure appears as a result of relatively weak molecular chirality (that is why it has a relatively large pitch), and the swirl direction of the spiral (left or right) is determined by the sign of the molecular chirality. Unlike this situation, the $N_{TB}$ nematics are formed as a result of spontaneous chirality breaking, they have nanoscale (a few nanometers) pitch.
This fact suggests that description of the twist-bend nematics in terms of an orientational elastic energy needs a modification related to relatively short pitch of the helicoidal structure. In the case the Frank moduli for the short-scale component of the director field are different from those for the long-scale component of the director. Therefore the components should be treated in terms of different elastic energies. We keep the notation $\bm n$ for the long-scale component of the director (nematic director in what follows) and introducing its short-scale component $\bm\varphi$. The components have to be orthogonal, $\bm n\cdot \bm \varphi=0$. Thus the vector $\bm\varphi$ has two independent components.
The quantity $\bm n+\bm\varphi$ can be naturally termed as the $N_{TB}$ phase  director. It has the helical conic structure in space. By other words, the short-scale component $\bm\varphi$ rotates around $\bm n$ at moving along the $\bm n$-direction. Therefore the absolute value of the vector $\bm\varphi$ gives the tilt angle $\theta $ for the conical spiral. Because the conical helical structure has a certain short pitch periodicity, it is characterized by the wave vector $q_0$, experimentally on the order of a few inverse molecular length.

Our paper is motivated by two very 
recent works \cite{KK20} - \cite{KK21} on rheological studies of the $N_{TB}$ liquid crystals.
The authors of these papers found nontrivial non-Newtonian behavior
of sheared $N_{TB}$ nematics. At relatively low shear rate (${\dot \gamma } \leq {\dot \gamma}_{c1}$) the stress
tensor $\sigma $ created by this shear strain, scales as $\sigma \propto {\dot \gamma }^{1/2}$. Thus the effective viscosity decreases with the shear rate ($\eta \propto {\dot \gamma }^{-1/2}$) manifesting so-called shear-thinning phenomenon. At intermediate shear rate ${\dot \gamma }_{c1} \leq {\dot \gamma} \leq {\dot \gamma }_{c2}$, $\sigma $ 
is almost independent of ${\dot \gamma }$ (a sort of plateau), and at large shear rate (${\dot \gamma } \geq 
{\dot \gamma}_{c2}$, $\sigma \propto {\dot \gamma }$, and it looks like Newtonian rheology. The critical values
of the shear rate (${\dot \gamma }_{c1}\, ,\, {\dot \gamma}_{c2}$) indicating transitions between dynamical regimes 
depend on temperature. Above certain temperature $T^*$ (below $N\, - \, N_{TB}$ phase transition point $T_c$, where $N$
stands for conventional nematic state)
the behavior becomes pure Newtonian. 
The aim of this paper is to present theoretical rationalization for the 
observed in these works \cite{KK20}, \cite{KK21} results. In what follows we integrate the input from recent works and discussions,
however my own contribution to this field will be also presented.

Although the origin of the observed experimentally rheological behavior of the $N_{TB}$ nematics still remains to be clarified,
the main message of this work is robust. Namely we claim that coarse grained dynamic description of the $N_{TB}$ phase is
allows to rationalize qualitatively the observed in the phase different rheological regimes. Such coarse grained description supplemented by
arguments based on the maximum rate of entropy production principle, can be used to estimate
the critical values of the shear rate separating these rheological regimes. The estimations made in the paper are applicable
to the shearing of well ordered samples of the $N_{TB}$ phase. If it is not the case, defects (disclinations or domain walls)
affect the rheology. However relying on the maximum entropy production rate principle, we expect that the rheological transitions
(crossover between different rheological regimes) are determined by the coarse grained effective viscosity coefficients. The latter ones
should be determined for the partially disordered (i.e., including defects) $N_{TB}$ phase flow.

{\bf{Basic derivation of the $N_{TB}$ coarse-grained theory.}}
${\, }$
The main feature which distinguishes the standard nematic $N$ and the twist-bend nematic $N_{TB}$ liquid 
crystals is a short
wavelength modulation of the orientation order ${\bm \varphi }$ presented in the $N_{TB}$ phase.
This two component vector ${\bm \varphi }$, orthogonal to the nematic director ${\bf n}$ (${\bm \varphi} \cdot  {\bf n}
= 0$) can be chosen as the order parameter describing $N$ - $N_{TB}$ phase transition. 
With this vector order parameter in hands one can write the Landau free energy functional. 

Taking into account the nature of the short-scale vector field $\bm\varphi$, we obtain (see more details in \cite{KL14},
\cite{KA17})
 \begin{eqnarray}
 \int dV \left\{\frac{a}{2} \bm\varphi^2
 +\frac{b_3}{8 q_0^2} \left[
 \left(n_i n_k \partial_i \partial_k +q_0^2\right) \bm\varphi \right]^2
 +\frac{b_1}{2} (\nabla \bm\varphi)^2
 \right. \nonumber \\ \left.
 +\frac{b_\perp}{2}\delta^\perp_{ij}
 \partial_i \bm \varphi \partial_j \bm \varphi
 +\frac{\lambda}{24} \varphi^4
 -\frac{\lambda_1}{16 q_0^2}
 \left(\epsilon_{ijk} \varphi_i \partial_j \varphi_k\right)^2
 \right\}, \quad
 \label{bana1}
 \end{eqnarray}
where $\delta^\perp_{ij}=\delta_{ij}-n_i n_j$. As usual, $a\propto T-T_c$, where $T_c$ is the mean field transition temperature. The quantities $b$ are analogs of the Frank moduli for the order parameter $\bm\varphi$. The free energy (\ref{bana1}) represents the minimal Landau model for the $N$--$N_{TB}$ phase transition,
neglecting fluctuations of the long-scale director $\bm n$, which anyway is suppressed under shear
(see e.g., \cite{GP93} - \cite{OP06}). Then denoting a preferred direction, $\bm n_0=(0,0,1)$
we represent the Landau functional (\ref{bana1}) in a more compact form by replacing the order parameter $\bm\varphi$ by its complex long-wavelength (!) counterpart $\bm\psi$
 \begin{equation}
 \bm\varphi= 2\,\mathrm{Re}\
 \left[\bm\psi \exp(i q_0 z) \right].
 \label{bana3}
 \end{equation}
If $\lambda_1>0$ then below the phase transition (at $a<0$) minimization of the free energy leads to the conical helical structure with
 \begin{equation}
 \varphi_x=2 |\psi_x|\cos(q_0z+\theta ), \quad
 \varphi_y=\pm 2|\psi_x|\sin(q_0z+\theta),
 \label{cone}
 \end{equation}
where $\theta $ is the phase of $\psi_x$ and signs $\pm$ correspond to two possible rotation directions of the conical structure.

It is worthwhile to mention the known in the literature other coarse grained descriptions of the $N_{TB}$
phase \cite{SK14}, \cite{PS16}, \cite{MD16} (I sincerely thank the anonymous reviewer for attracting my attention to these papers).
Although the approaches \cite{SK14}, \cite{PS16}, \cite{MD16} are conceptually similar 
to the model presented in this work, there is an essential difference. In our approach the long wavelength (nematic director) and
short wavelength the $N_{TB}$ order parameter are explicitly separated into the Landau free energy expansion (\ref{bana1}). Technically
the short wavelength nature of the $N_{TB}$ order parameter yields to the specific form of the gradient terms, providing the
softening of the order parameter in the vicinity of a circle in the reciprocal space (not around a single point (zero wave vector) 
as it is the case for the nematic order parameter). In the $N_{TB}$ phase but not too far from the transition
point $T_c$ (that is always the case due to $(T - T_c)/T_c \ll 1$)
the entering Landau theory (\ref{bana1}) parameters $b_3$, $b_1$, $b_\perp$, $\lambda $, $\lambda_1$ and $q_0$ can be considered in the mean-field
approximation as temperature independent, and only the controlling mean-field behavior coefficient $a$ scales
as $T - T_c$. Soft near the circle $|{\bf q}| = q_0$ in the reciprocal space fluctuations of the $N_{TB}$ order parameter
changes this mean-field behavior (see more details below and also in \cite{KL14}).

Since at the scales $r$ larger than $q_0^{-1}$ we are interested in this work, the $N_{TB}$ phase free energy
and dynamic equations coincide with those for the effective smectic $A$, in what follows we will use interchangeably the both terms. 
$N_{TB}$ nematics when discussing generic features of the $N \to N_{TB}$ phase transition, and effective smectic $A$, speaking about 
rheology.

Fluctuations around this mean-field state include long wavelength fluctuations
\begin{equation}
{\bf n} = {\bf n}_0 + \delta {\bf n},
 \label{kats1}
 \end{equation}
and short wavelength fluctuations ${\bm \varphi }$ which is orthogonal to ${\bf n}$
\begin{equation}
{\bm {\varphi }_\perp} = 2 Re \left [\bm {\psi }e^{iq_0z}\right ]\, ;\,
{\bm {\varphi }_{||}} = -2\delta {\bf n} Re\left [\bm {\psi }e^{iq_0z}\right ] .
 \label{kats2}
 \end{equation}

Two remarks are in order here. First, as it was shown previously (\cite{BR92} -\cite{MF02} externally imposed shear flow 
suppresses long wavelength fluctuations. Second, renormalization group flow draws the system towards a symmetric state \cite{WK74}, 
therefore, the renormalized value of $b_1$ coefficient tends to zero.
Then, the proposed (\ref{bana1}) model is equivalent to the standard De Gennes model \cite{GP93} describing nematic - smectic $A$
phase transition, \cite{GP93}, \cite{KL93}-\cite{CL00}. It is worth to noting, that there is no density modulation in the liquid-like
$N_{TB}$ phase. However, conical $N_{TB}$ orientation order, on scales much larger than the orientation 
modulation period looks like as a periodic in space layered smectic-like structure.

A wide range of coarse-grained models have been proposed, usually dedicated to modeling of multiscale systems.
Coarse graining allows to decrease a number of essential degrees of freedom at the expense of microscopic details
(e.g, replacing individual building blocks by their groups). Technically in this work we use slightly modified to be applicable
to the $N_{TB}$ phase coarse-graining procedure described in \cite{GP93} for the static behavior of cholesterics,
and in \cite{KL86} - for dynamics of cholesteric liquid crystals. 
The properties of the deformed $N_{TB}$ liquid crystal depends essentially on the ratio of the inhomogeneity
scale (say $r_{in}$) and the $N_{TB}$ heliconical spiral pitch $q_0^{-1}$. At scales $r_{in} \gg q_0^{-1}$ 
to find large-scale static and dynamic characteristics, we have to eliminate fast degrees of freedom. It is done
\cite{GP93}, \cite{KL86}
by appropriate integration out the fast degrees of freedom. Performing the averaging we express the coarse grained
coefficients (elastic moduli and viscosity coefficients) in terms of the bare $N_{TB}$ parameters defined at scales
$\ll q_0^{-1}$.

The elastic energy of this coarse grained phase
reads as
\begin{equation}
{\tilde {F}} = \frac{1}{2}\int d^3 r\left [ {\tilde B}\left (\frac{\partial u}{\partial z}\right )^2 + {\tilde K}\left 
(\nabla _\perp ^2 u\right )^2\right ]
.
 \label{kats3}
 \end{equation}
Here $u$ is an orientation layer displacement along the normal to the layer ($z$ axis in our notation), and coarse-grained
elastic moduli \cite{GP93}, \cite{KL93} are 
\begin{equation}
{\tilde B} = b_\perp q_0^2\, ;\, {\tilde K} = \frac{3}{8} b_\perp
.
 \label{kats4}
 \end{equation}

{\bf{Coarse-grained rheology of the $N_{TB}$ phase.}}
${\, }$
\begin{figure}
\begin{center}
\includegraphics[height=2in]{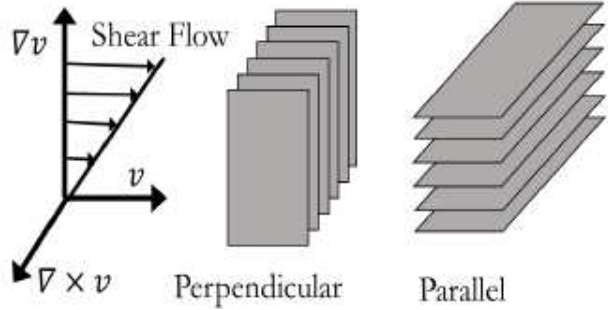}
\end{center}
\caption{Two rheological configurations considered in the paper.}
\label{f1}
\end{figure}
Since we are interested in rheological behavior (dynamics) the equilibrium elastic energy (\ref{kats3}) should be supplemented by the coarse grained
viscous (dissipative) stress tensor (see its derivation in \cite{KL86}, \cite{KL93}). In what follows
we consider two main rheological configurations (see Figure 1) for the effective smectic, representing $N_{TB}$
nematic at large scales. 
Namely, perpendicular configuration (where the normal to the layer, unit vector ${\bf l}$,
perpendicular to the
imposed shear velocity circulation $curl {\bf v}$) and parallel configuration,
where ${\bf l}$ parallel to $curl {\bf v}$. 

It is worth to note that in the both configurations (${\bf l}$ parallel to $curl {\bf v}$,
or ${\bf l}$,
perpendicular to the
$curl {\bf v}$) the shear velocity ${\bf v}$
is always oriented along the layers of the effective smectic.

Coarse grained (smectic-like 5 viscosity coefficients [37] are expressed in terms of the bare (nematic-like)
According to the definitions [27], [35] the uniaxial nematic phase $N$ is characterized
5 independent viscosity coefficients $\eta _1 \, -\, \eta _5$ (or 6 Leslie's
viscosity coefficients, which have to satisfied one constraint \cite{GP93}). At small space scales ($r < q_0^{-1}$) these
viscosity coefficients are certainly different in the $N$ and $N_{TB}$ phases. However the difference is relatively small 
(on the order of $\bm\varphi ^2$),
and we neglect the difference in what follows. What is relevant for us is the fact that for two rheological configurations 
(see figure 1)
we are interested in this work, only two combinations of the five viscosity coefficients needed for a full description of dynamical behavior.
The flow is determined by these two combinations of the coarse grained viscosity coefficients.
Namely, ${\tilde \eta }_2$ coefficient determines the shear flow rheology for the perpendicular configuration,
and the coarse grained viscosity ${\tilde
\eta }_3$ determines the flow in the parallel configuration. It is worth to stress that coarse grained
free energy of the $N_{TB}$ phase coincides with the smectic $A$ free energy, provided the smectic layer displacement $u$
is replaced by the phase of the $N_{TB}$ order parameter.
Similar statements holds for the coarse grained dynamics of the $N_{TB}$ phase. However we have to keep in mind that 
the coarse grained viscosity coefficients have a physical meaning only at the scales $r$ larger than $q_0^{-1}$. At smaller
scales rheological behavior is determined by the bare viscosity coefficients. Therefore when the shearing liquid crystal sample
thickness becomes smaller than $q_0^{-1}$ the coarse grained approximation is meaningless. In particularly no any room for the coarse
grained theory in the $N$ phase, where the $N_{TB}$ order parameter $|\bm \varphi | \to 0$ and hence $q_0 \equiv 0$.

In the rheological configurations shown in figure \ref{f1} these two coarse grained viscosity coefficients can be expressed \cite{KL86} in terms of the bare 
$N_{TB}$ viscosity coefficients
($\eta _i$, with $i = 1,...5$) as
\begin{equation}
{\tilde \eta}_2 = 2\eta _2 - \frac{1}{2} \eta _3 + \frac{1}{4}\eta _1\, ;\, 
{\tilde \eta}_3 = \frac{1}{4}\left [2\eta _3 + 2\eta _2 - \eta _1\right ]
.
 \label{kats5}
 \end{equation}
For typical in nematic liquid crystals values of the bare viscosity
coefficients \cite{GP93}, \cite{KL06}, \cite{OP06}, we estimate from (\ref{kats5})
the values of the needed for us coarse grained bare
viscosity coefficients: 
\begin{equation}
{\tilde \eta}_2^0 \simeq 0.832\, Poise\, ;\, 
{\tilde \eta}_3^0 \simeq 0.2\, Poise
.
 \label{kats6}
 \end{equation}

Because in the limit ${\dot{\gamma}} \to 0 $, ${\tilde \eta }_2 > {\tilde \eta}_3$ just the perpendicular configuration is preferable
one (the configuration leads to the maximum rate for the entropy production). 

This principle (the maximum entropy production rate) as well as the apparently controversial statement (the principle of minimum entropy production principle) in each particular condition should be reconsidered, based on statistical mechanics,
and hydrodynamics to find a stationary state for non-equilibrium systems. In this work we rely on the mapping of 
coarse grained dynamics of the $N_{TB}$ phase into that for the smectic $A$ liquid crystals.
With this mapping in hands the three regimes observed in the sheared $N_{TB}$ liquid crystals can be rationalized 
similarly to the known for smectic $A$ results (see e.g., two review articles \cite{SB10}, \cite{FK14}), and the results
suggest three rheological regimes which are governed by the maximum entropy production rate (irrespective to physical
mechanisms behind).

To find the rheological relation for
such configuration, and then the rheological phase diagram, one has to analyze the effective smectic order parameter
fluctuations, and how the fluctuations are affected by the imposed shear ${\dot{\gamma}}$.
Our approach is more intuitive and qualitative than just solving the dynamical equations with imposed shear
\cite{BR92} - \cite{MF02}. However we expect that it will be useful, offering a deeper insight into physics behind
the $N_{TB}$ phase rheology. 

Our theory predictions are based on the following observations. The shear viscosity in the
perpendicular configuration ${\tilde \eta}_2$ scales
proportional to the correlation length \cite{OP06}
\begin{equation}
{\tilde \eta}_2 \propto \xi 
.
 \label{kats7} 
\end{equation}
To proceed further with the rheological relation we have to find how $\xi$ scales with the shear rate
${\dot \gamma}$. According to the Landau - Brazovskii weak crystallization theory (see the original paper \cite{BR75}, 
on thermodynamics and review article \cite{KLM93}, which includes also dynamics of the weak-crystallization
transition), the correlation length scales as 
\begin{equation}
\xi \propto \Delta ^{1/2} 
,
 \label{kats8} 
\end{equation}
where $\Delta $ is the gap in the soft mode, describing weak first order transition 
with emerging one, two, or three
dimensional translational ordering. Said above leads to the following
static correlation function for the order parameter \cite{BR75}
\begin{equation}
\frac{T}{\Delta + \alpha (q-q_0)^2} 
,
 \label{kats9} 
\end{equation}
where $\alpha $ is a phenomenological coefficient translated into the elastic modulus
below phase transition point. This form (\ref{kats9}) of the correlation function is
a specific feature of the weak crystallization transitions, where the critical
mode softening occurs at the finite wave vector $q_0$. In dynamics \cite{KLM93}
the dynamic response function corresponding to the presented above (\ref{kats9})
correlation function reads as
\begin{equation}
\frac{T}{\omega - i [\Delta + \alpha (q-q_0)^2 ]}
.
 \label{kats10} 
\end{equation}
We see from (\ref{kats10}) that $\Delta $ scales as the frequency $\omega $, and therefore
in the shear rheology condition as the shear rate ${\dot \gamma }$.
Finally combining everything together (\ref{kats8}) - (\ref{kats10}) we conclude that in a low shear rate
\begin{equation}
\tilde {\eta }_2 \propto ({\dot \gamma})^{-1/2}
 \label{kats11} 
\end{equation}
in the agreement with experimental data \cite{KK20} - \cite{KK21}.
Now we are in the position to estimate the first threshold value ${\dot \gamma }_{c1}$ for the shear rate.
This threshold corresponds to the condition when ${\tilde \eta }_2$ becomes on the order of the bare value
of ${\tilde \eta}_3^0$:
\begin{equation}
{\dot \gamma }_{c1} \propto \left (\frac{\tilde {\eta }_2^0}{\tilde {\eta }_3^0}\right )^2
 \label{kats12} 
\end{equation} 
At the shear rate larger than the threshold ${\dot \gamma } >
{\dot \gamma }_{c1}$ a part of the $N_{TB}$ sample converts into the parallel configuration. Then, similar to
the famous Van-der-Waals - Maxwell construction \cite{LL80}, \cite{HU87}, the rheological curve
corresponds to a sort of plateau. The plateau holds until when the entire volume of the $N_{TB}$ 
phase transforms into the parallel
configuration. In the parallel configuration according to \cite{OP06} ${\tilde \eta }_3
\propto \xi ^{5/2}$, and therefore the weak crystallization theory \cite{KLM93} predicts 
${\tilde \eta }_3 \propto ({\dot {\gamma }})^{-5/4}$. With this scaling in hands
we estimate the second shear rate threshold ${\dot \gamma }_{c2}$
as
\begin{equation}
{\dot \gamma }_{c2} \propto \left (\frac{\tilde {\eta }_2^0}{\tilde {\eta }_3^0}\right )^4
 \label{kats13} 
.
\end{equation} 
Finally when the $N_{TB}$ sample transforms completely into the parallel configuration (i.e.,
at  ${\dot \gamma } > {\dot \gamma }_{c2}$) we face to the situation discussed first long ago by De Gennes
\cite{GE76} (see also more details concerning smectics in the works \cite{BR92} - \cite{MF02}).
Namely, a large shear rate ${\dot \gamma } > {\dot \gamma }_{c2}$ suppresses the order parameter
fluctuations. Then we expect ${\tilde \eta} _3$ is shear rate independent quantity and we
have the Newtonian-like rheology. Thus our simple model qualitatively explains all three
rheological regimes observed experimentally \cite{KK20} - \cite{KK21} in the $N_{TB}$ phase.

{\bf{Outlook and Conclusions.}}
${\, }$

Recent progress in rheology of the $N_{TB}$ liquid crystals has led to a number of new and exciting experimental results 
\cite{KK20}, \cite{KK21}. In the paper we propose a simple heuristic approach to rationalize these new experimental data.

The approach is based on the coarse grained dynamic description of the $N_{TB}$ phase, valid for the space scales
larger than the pitch of the $N_{TB}$ heliconical structure. Such description supplemented by the semi-qualitative arguments, relayed
on the maximum entropy production rate, is used to estimate critical shear rate magnitudes separating different rheological regimes.
The corresponding thresholds are determined by the coarse grained viscosities of the system.
In own turn the coarse grained parameters entering the theory
are expressed in terms of local quantities of the bare nematic liquid crystal. The latter characteristics could be measured
by X-ray, NMR or other basically local probes.

The key starting point of our approach is based on a simple observation that the anisotropic viscous properties of the liquid crystals
introduce a host of novel phenomena in rheology.  We find that at relatively low shear rate (${\dot \gamma } \leq {\dot \gamma}_{c1}$) the stress
tensor $\sigma $ created by this shear strain, scales as $\sigma \propto {\dot \gamma }^{1/2}$. Thus the effective viscosity decreases with the shear rate ($\eta \propto {\dot \gamma }^{-1/2}$) manifesting so-called shear-thinning phenomenon. At intermediate shear rate ${\dot \gamma }_{c1} \leq {\dot \gamma} \leq {\dot \gamma }_{c2}$, $\sigma $ 
is almost independent of ${\dot \gamma }$ (a sort of plateau), and at large shear rate (${\dot \gamma } \geq 
{\dot \gamma}_{c2}$), $\sigma \propto {\dot \gamma }$, and it looks like Newtonian rheology. Within our theory
the critical values
of the shear rate scales as ${\dot \gamma }_{c1}\propto ({\tilde {\eta }_2^0}/{\tilde {\eta }_3^0})^2$,
and ${\dot \gamma}_{c2} \propto ({\tilde {\eta }_2^0}/{\tilde {\eta }_3^0})^4$ respectively. Here
${\tilde \eta }_2^0$ and ${\tilde \eta }_3^0$ are bare coarse grained shear viscosity coefficients 
of the effective smectics equivalent to the $N_{TB}$ phase at large scales. Our mainly qualitative theory may not have the right
numbers for the dynamic shear rate thresholds. However theory predicts the right scaling laws observed in the experiments.
Our consideration suggests that the described phenomena and mechanisms can bring about different rheological scenarios worthy of 
further studies. In this work we have only scratched the surface of this reach subject, focusing only on the most simple
questions,
which can be answered by calculations ''on a back of the envelope''.

The main message of this work is that coarse grained dynamic description of the $N_{TB}$ phase
allows to rationalized observed in the phase different rheological regimes. Such coarse grained description supplemented by
qualitative arguments based on the maximum rate of entropy production principle, can be used to estimate
the critical values of the shear rate separating these rheological regimes. The estimations made in the paper are applicable
to the shearing of well ordered samples of the $N_{TB}$ phase. If it is not the case, defects (disclinations or domain walls)
affect the rheology. However relying on the maximum entropy production rate principle, we expect that the rheological transitions
(crossover between different rheological regimes) are determined by the coarse grained effective viscosity coefficients. The latter ones
should be determined for the partially disordered (i.e., including defects) $N_{TB}$ phase flow. 
Guided by the modest aim of this work, we present only scaling analysis of  three regimes of a steady shear  viscosity
curve which is in qualitative agreement with previously reported observations \cite{KK20}, \cite{KK21}, \cite{PS16}.
Even more not all of the observed qualitative features of the rheological curves are reproduced in our approach.
For example there is no independent of shear rate plateau in the interval ${\dot \gamma }_{c1} \leq {\dot \gamma} \leq {\dot \gamma }_{c2}$.
Instead of that $\sigma \propto {\dot \gamma }^{\alpha }$, where the exponent $\alpha \simeq 0.1$ is small but not zero.
The matter is that the estimated thresholds of rheological curves assume a
well-defined orientation of the director respective to the shear flow. Most probably it is not the case
in mentioned above experimental works, especially in the regions of relatively small shear rates. However, 
the ordered, defects-free states can be achieved, e.g., by orienting liquid crystal external
field. Such investigation is beyond the scope of this work, direct experimental measurements of the ordered $N_{TB}$ phase
rheology is still a challenging task. I hope that this paper will stimulate
discussions on the intriguing and important issues of non-Newtonian rheology in liquid crystals.

\acknowledgements 

I am dedicating this work  to outstanding theoretical
physicist and remarkable person Mark Azbel (12.05.1932\, -\, 31.03.2020). For me personally it was a rewarding pleasure to start my work at the Landau Institute in Chernogolovka at the time when M.Azbel was amongst the permanent staff of the Institute. At that time my understanding of 
theoretical physics, benefited tremendously from listening discussions of the ''grands''. One of the sad consequences of the very fast
developments of physics is that it looses its unity. Mark Azbel was one of a few Landau Institute scientists who worked
in a very broad field, including soft matter physics, the subject of this manuscript.

\end{document}